# Rare-Event Estimation for Dynamic Fault Trees

SERGEY POROTSKY

**Abstract**. Article describes the results of the development and using of Rare-Event Monte-Carlo Simulation Algorithms for Dynamic Fault Trees Estimation. For Fault Trees estimation usually analytical methods are used (Minimal Cut sets, Markov Chains, etc.), but for complex models with Dynamic Gates it is necessary to use Monte-Carlo simulation with combination of Importance Sampling method. Proposed article describes approach for this problem solution according for specific features of Dynamic Fault Trees. There are assumed, that failures are non-repairable with general distribution functions of times to failures (there may be Exponential distribution, Weibull, Normal and Log-Normal, etc.). Expessions for Importance Sampling Re-Calculations are proposed and some numerical results are considered.

## 1. INTRODUCTION

One of the important tasks of the Reliability Estimation is Analysis of the Fault Tree. Building and calculation of the Fault Tree are considered in the [1 - 3]. Usually analytical methods are used (Minimal Cut sets, Markov Chains, etc.), but sometimes, for complex models, it is necessary to use Monte-Carlo simulation. A problem of Fault Trees calculation is considered one of the most complex ones, since structure of such trees is characterized by a considerable number of interconnections.

Fault Trees with Dynamic Gates are often used in some specific fields of reliability. Examples of such gates are PAND (Priority AND), SEQ (Sequence Enforcing), SPARE, etc. Classical Fault Tree Analysis methods (Minimal Cut Sets calculations) are applicable only for Static Fault Trees. Using of analytical methods, based on Markov Chain methods, are restricted only for dynamic trees with very low scalability. For large Fault Trees may be used approximate method, proposed on the SAE ARP 4761 [1] – to calculate probability of required order of failures and to use calculated value as additional event. Unfortunately, this approach was developed only for PAND gate and isn't applicable for other types of Dynamic Gates (SEQ, SPARE, etc.). Moreover, even for PAND gate this approach get us only very and very approximate estimations and hard applicable for Fault Trees, which have Basic Events with different Mean Values of Time to Failure (MTTF). In general case the Monte Carlo method is used [4–6, 9–13].

Usually reliability estimation has high requirements for **Probability** – for example, it has to be less than $10^{-8}...10^{-10}$; so, it will be rare event. Estimation of rare-event probability by means of the direct Monte-Carlo method is impossible, because it requires a lot of simulation cycles (at least $10^{9}...10^{11}$). Standard way to reduce computational time and to improve the simulation accuracy is the Variance Reduction technique (Importance Sampling) [7 - 9].
For rare-event estimation the Importance Sampling method is used and most essential problem on this method – how to select appropriate reference probability distribution. Unfortunately, well-known approaches (e.g., [7 - 9]) for reference distribution selection (scaling, translation) are not applicable for Dynamic Fault Trees analysis. Reason is following – classical rare-event estimation task allows to calculate

Probabilty{ $S(x_1,…,x_N) > T$ } for very large T, by means of Importance Sampling method using. It is assumes, that function S "is good in some sense", e.g. it is combination of Min, Max, Sum, etc. Typical example of the "good function" S is shortest path calculation. For Fault Tree rare-event estimation the task is some another – to calculate Probabilty{ $H(x_1,…,x_N) < T$ }, where H is failure time of the fault tree TOP, $x_1,…,x_N$ are failure times of the basic events 1…N, and T is mission time. Certainly, it is possible to transform this task for the classical task by means of estimation of Probabilty{ $1/H(x_1,…,x_N) > 1/T$ }, but in this formulation the function $S(x_1,…,x_N) = 1/H(x_1,…,x_N)$ will not be "good" as supposed for the classical task and so results of rare-event estimation will be non-correct.

Proposed article describes approach for this problem solution according for specific features of Dynamic Fault Trees. Some single aspects of this problem are considered in different articles, denoted for Fault Tree Monte Carlo simulation. For example [4, 10] consider using of Monte Carlo simulation for Dynamic Fault Tree Analysis, but they use direct simulation, so they are not applicable for rare event simulation.

Articles [5, 9] propose to use Importance Sampling for estimate TOP probability of Fault Trees. But suggested formulas don't allow take into account order of events, so they are not applicable for dynamic fault trees. Article [6] considers Importance Sampling using for Dynamic Fault Trees, but suggested formulas (as on [5]) correspond only for Static Fault Tree, because they don't take into account order of events.

## 2. ALGORITHM DESCRIPTION

:Table of main definitions is below

| | |
|---|---|
| T | Length of Sysstem Life |
| N | Amount of Basic events |
| i | Index of Basic Event ( i = 1…N) |
| $x_i$ | Failure Time of i-th Basic Event |
| K | Amount of Cycles to perform Main Simulation(for default = 100,000) |
| j | Index of Simulation Cycle ( j = 1…K) |
| $f_i(t)$ | Probability Density Function (PDF) of i-th Basic Event Failure Time |
| $F_i(t)$ | Cumulative distribution Function (CDF) of i-th Basic Event Failure Time |
| $g_i(t)$ | Reference Probability Density Function of i-th Basic Event Failure Time |
| $G_i(t)$ | Reference Cumulative Distribution function of i-th Basic Event Failure Time |
| P | Probability of TOP Failure |
| K_Prelim | Amount of Cycles to perform Preliminary Simulation (for default = 1000) |
| $v_i$ | Reference Parameter for Reference Probability Density function $g_i(t)$ of i-th Basic Event |

| D | Common (for all Basic Events) Secondary Reference Parameter |
|---|---|
| AmPos | Amount of simulation cycles, for which TOP = Failure |
| AmPos_Up | Upper Bound of AmPos for Preliminary Simulation (for default = 100) |
| AmPos_Dn | Down Bound of Am_Pos for Preliminary Simulation (for default = 10) |
| IC | Iteration Counter for step-by-step Preliminary Simulation |
| D_Up | Current Upper Bound of D value |
| D_Dn | Current Down Bound of D value |

A Fault Tree is a Directed Acyclic Graph in which the leaves are basic events and the other elements are gates. Using Boolean Algebra Laws, usually any static Fault Tree may be represented by means of two types of gates: *AND gate* which fails if all inputs fail; *OR gate* which fails if at least one of its inputs fails. Other, more complex gates (e.g., "K out of M" gate, named as Voting gate), may be expressed as combination of *AND gates* and *OR gates*. Assume, that inputs for some gate are characterized by the failure times of $z_1,\ldots,z_q$ – there may be Basic Events or outputs of some intermediate gates; let us y is failure time of gate output. During Fault Tree Monte-Carlo simulation we use following formulas:

- For gate OR : $y = \min\{z_1,\ldots,z_q\}$ (1)
- For gate AND : $y = \max\{z_1,\ldots,z_q\}$ (2)

On the static Fault Tree the value of "TOP = Failure" really depends only of Boolean states of the Basic Events (True Versus False), i.e. really don't depend on failure times of Basic Events, rather on condition, when this failure was – before of after time T.

Dynamic Fault Trees extends static Fault Trees with the following dynamic gates:
- Priority AND gate (**PAND**) gate is a gate which fails when all its inputs fail from left to right in order. For this gate $y = z_q$, if $z_1 \leq z_2 \leq \ldots \leq z_q$; else y = Infinite. (3)
- Sequence Enforcing gate(**SEQ**), for which $y = z_1 + z_2 + \ldots + z_q$ (4)
- **SPARE** gate, for which $y = z_1$, if $z_2 < a*z_1$ else
$y = (1-a)*z_1 + z_2$. In this formula "a" parameter is the dormancy factor of the second input. (5)

Consider Fault Tree with Basic Events, for which the failures are non-repairable with Probability Density Function (PDF) $f_i(t)$ and Cumulative Distribution Function (CDF) $F_i(t)$; the corresponding probability, that failure of Basic Event i will be before time T, is $p_i = F_i(T)$. Our task is to estimate the probability of "TOP = Failure":

P = Probability{TOP = Failure}. For very small values of $p_i$ this rare event estimation needs a very large number of simulations. Importance Sampling approach get us possibility to deal with new probabilities $q_i$ instead of real values $p_i$ and so the main problem is to select optimal values for $q_i$ values [5, 6, 9]. These approaches are applicable only for Static Fault Trees, i.e. "**TOP = Failure**" is independent of different failures times of Basic Events – important only, these times less or more than time T. For Dynamic Fault Trees the Important Sampling should use values of PDF functions $f_i(t)$ instead of probability values $p_i$. Define $g_i(t)$ and $G_i(t)$ – new (reference) probability density and cumulative distribution functions for failure time of Basic Event i. Value of P will be following:

$$P = \frac{1}{K} * \sum_{j=1}^{K} I(t_j) * \left[ \frac{\prod_{i=1}^{N} f_i(t_{j_i})}{\prod_{i=1}^{N} g_i(t_{j_i})} \right] \text{, where}$$

K – Amount of simulation cycles, j – index of simulation cycle (j = 1…K)

N – Amount of Basic Events, i – index of Basic Event (i = 1…N)

**$t_j$** = $(t_{j1},…t_{ji},……,t_{jN})$ - vector of Basic Event failure times for simulation cycle number j with reference vector probability density function **g**(t), j = 1…K

I(**$t_j$**) – indicator function for simulation number j, I(**$t_j$**)=1, if S(**$t_j$**) < T; otherwise, I(**$t_j$**)=0.

S(**t**) – function to calculate TOP according Fault Tree structure for vector **t** of Basic Event failure times

However, the Fault Tree rare-event estimation technique, based on these expressions, cannot always guarantee the successful results. For example, above formula for TOP probability P doesn't get us correct solution even for simplest case – exponential PDF function $f_i(t)$ for all Basic Event failure times and simplest static gate OR. Such a situation is typical for a Fault Trees, in which some gates are OR and some gates are PAND. To get correct solution, it is necessaryto to modify Importance Sampling expression. If for some Basic Event i the failure time will be more than time T, it isn't significant for event "TOP = Failure" – in what concrete time was failure of the Basic Event i. It is understood, that Dynamic Fault Tree should really use concrete values of Basic Event failure times, there are not enough to use only Boolean values (less than T or more than T) – but only for values less than T!

It has been proposed and proven, that for gates **OR, AND, PAND, SEQ** and **SPARE:**

If the failure time **y** of the gate output is less than timr T, then it independent of concrete values $x_i$ of gate inputs, for which $x_i \geq T$ – significant are only concrete values $x_i$, for which $x_i < T$ and boolean values (false) for which $x_i \geq T$.

This statement allows modifying the equation for **Probability**{TOP = Failure}:

$$P = \frac{1}{K} * \sum_{j=1}^{K} I(t_j) * \left[ \frac{\prod_{i=1}^{N} f\_modif_i(t_{j_i})}{\prod_{i=1}^{N} g\_modif_i(t_{j_i})} \right] \text{, where} \qquad (6)$$

f_modif$_i$ (t$_{ji}$) = f$_i$(t$_{ji}$), if  t$_{ji}$<T, otherwise f_modif$_i$(t$_{ji}$) = 1 – F$_i$(T)  (7)

g_modif$_i$(t$_{ji}$) = g$_i$(t$_{ji}$), if  t$_{ji}$<T, otherwise  g_modif$_i$(t$_{ji}$) = 1 – G$_i$(T)  (8)

Such it is necessary to use Mixed Continuous-Discrete PDFs (both for initial f(t) and reference g(t) ) instead of usually used pure Continuous PDFs.
Based of above proposed modified Importance Sampling equation, were proposed the original procedure to select the optimal values of reference probability density functions g$_i$(t). Reference probability density functions g$_i$(t) selection for each Basic Event i is started by building an initial type of g$_i$(t). Although there are many kinds of possible transformations, the following two approaches are most widely used for Importance Sampling: Scaling and Translation.

For Scaling using we define $g(t) = \frac{1}{a} * f(t/a)$. For example, if f(t) is Exponential PDF with $f(t) = \frac{1}{u} * \exp(-t/u)$, we will get, that $g(t) = \frac{1}{v} * \exp(-t/v)$ and, so, $G(t) = 1 - \exp(-t/v)$ where v is control (unknown and has to be defined) reference parameter.

If f(t) is Weibulll PDF with f(t) = b*(u$^{(-b)}$)*(t$^{(b-1)}$)*exp(-(t/u)$^b$), we will get, that g(t) = b*(v$^{(-b)}$)*(t$^{(b-1)}$)*exp(-(t/v)$^b$),  and, so, G(t) = 1 - exp(-(t/v)$^b$), where v is reference control parameter.

For Translation using we define g(t) = f(t - a), where "a" is control parameter and has to be chosen. Other Importance Sampling Transformation also may be used. For our point of view, for Dynamic Fault Tree rare-event estimation the best solution is to use Scaling transformation.
After the type of g$_i$(t) is selected for each i-th Basic Event (i = 1…N), it is necessary to choice the optimal value of the control parameter v$_i$. It is performed by means of Monte-Carlo simulation of evaluated Failt Tree with small amount K_Prelim of simulation cycles (usually it is enough to use K_Prelim = 1000). For current simulation cycle first there are calculated failure times for each of the Basic Events - according early built reference probability density functions g_modif$_i$(t) with some control parameter v$_i$. For each Basic Event i it is generated a random value x$_i$. These values are propagated through the fault tree according gates and formulas (1)…(5). This is done until the TOP node is arrived at which point a sampled failure time of the entire tree is calculated. After this the amount of simulation cycles, for which "TOP = Failure" time less than time T, is calculated value of AmPos.

For each of the Basic Event number i the following equations are proposed to calculate values of v$_i$:
- $1 - G_i(t) = \frac{1 - F_i(t)}{D}$  (9)

where D is some common (for all Basic Events) secondary control reference parameter. For using of Importance Sampling Scaling transformation the following expressions are proved:

- $v_i = 1 / \left( \dfrac{1}{u_i} + \dfrac{\log(D)}{T} \right)$ - for Exponential PDF of $f_i(t)$     (10)

- $v_i = \dfrac{T}{\left( \left( \dfrac{T}{u_i} \right)^{b_i} + \log(D) \right)^{\frac{1}{b_i}}}$ - for Weibull PDF of $f_i(t)$     (11)

Also may be used several secondary control reference parameters $D_1,\ldots, D_s$. To define values of primary control reference parameters $v_1,\ldots,v_i,\ldots v_N$ some other expressions may be used.

For first iteration (IC = 1) the reference parameters $v_i$ are setted equaled for $u_i$ for all Basic Events 1…N (it corresponds D = 1). It is necessary to perform Fault Tree Monte-Carlo preliminary simulation (according PDF functions g_modif$_i$(t) of Basic Events and formulas (1)…(5) ). After this the Amount of simulation cycles (from full amount of simulation cycles, equaled to K_Prelim), for which time of the {TOP = Failure} less than timeT, is calculated - it is Am_Pos. If after first preliminary simulation with K_Prelim cycles we get AmPos[IC=1] > 0, Importance Sampling isn't required and it is necessary simply to continue simulation up K simulation cycles. If AmPos[IC=1] == 0, it is necessary to choice $v_i$ values. Following main schema to define optimal values of $v_i$ is proposed:

*If* AmPos_Dn ≤ AmPos ≤ AmPos_Up, the current values $v_i$ are selected as optimal *else* it is necessary to change value of D according received values of AmPos on previous simulation steps, to increment Iteration Counter (IC) and to repeat Monte-Carlo simulation of the Fault Tree with new $v_i$ values (due to new D value, according formulas (9)…(11)) and same sample size of K_Prelim simulation cycles. For default the "tuned" values are settled as: K_Prelim = 1000, AmPos_Dn = 10, AmPos_Up = 100. Details of proposed procedure of D changing, based on method of secants, are presented on fig. 1. After the optimal values of the reference parameters $v_i$ are calculated, it is performed the final Monte-Carlo simulation of evaluated Dynamic Failt Tree with amount K of simulation cycles (usually it is enough to use K = 100,000). Calculation of the value of P is performed according formulas (6)…(8).

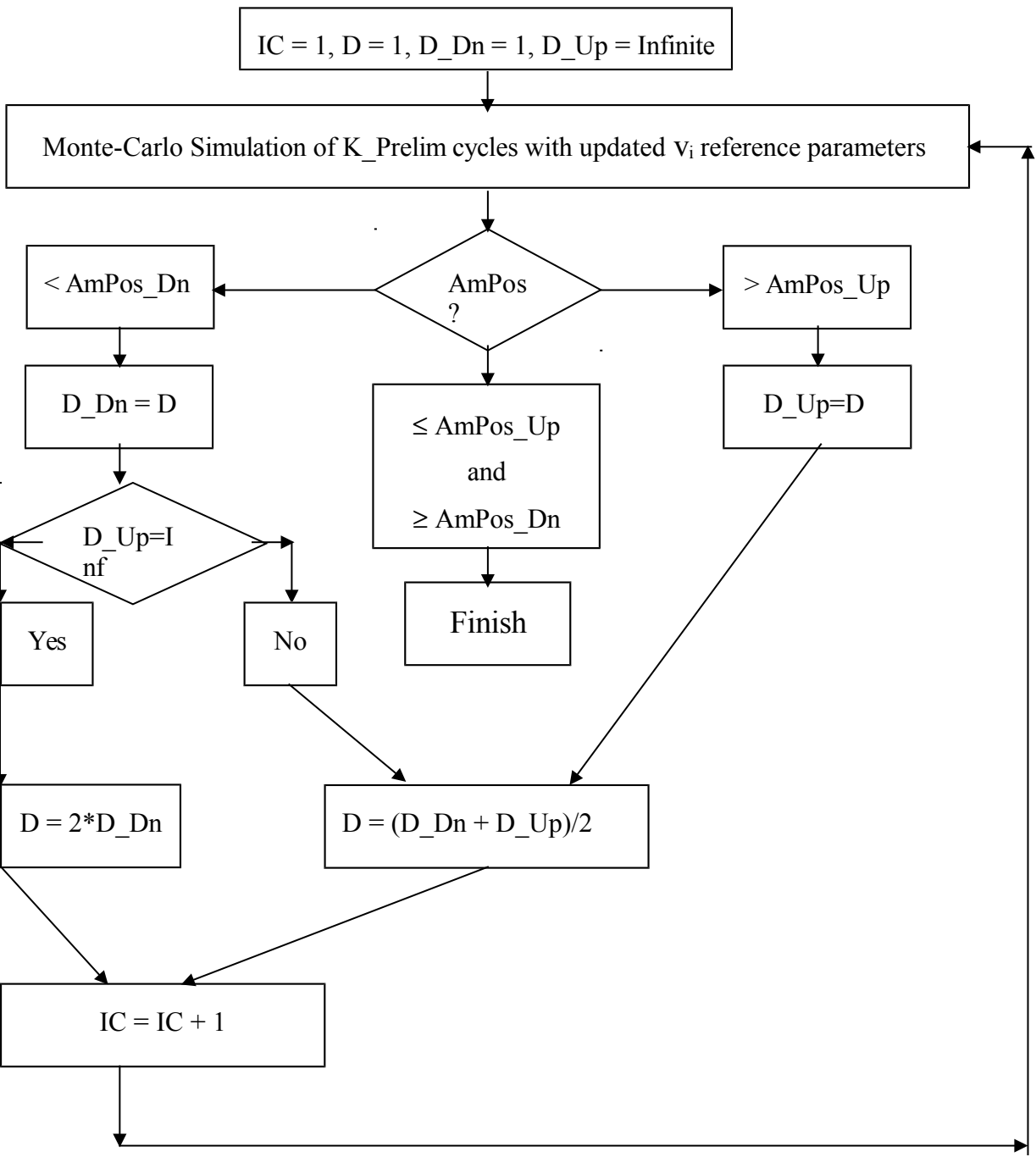

FIG. 1. Schematic flowchart to select optimal values of reference parameter.

## 3. NUMERICAL EXAMPLE

Consider Fault Tree with following parameters:

$T = 1$, $N = 4$, $F_i(t) = 1 - \exp(-T/u_i))$, where $u_i = 1000*i$, $i = 1…N$.

Structure of the Fault Tree is following:

TOP = (BE$_1$ **AND** BE$_2$ **AND** BE$_3$) **PAND** (BE$_2$ **AND** BE$_3$ **AND** BE$_4$),

where BE$_i$ is Basic Event with index i.

It is seen, that this Fault Tree has strong overlap between two parts – each of the part containts 3 BEs, and 2 BEs of them are common for two parts.

To select optimal value of the Secondary Reference Parameter D it was performed the Monte-Carlo simulation of 1000 cycles (i.e. K_Prelim = 1000).

Table below illustrates proposed method explaining a quick way of finding the optimal values of reference parameters.

| INPUT | | | | OUTPUT |
|---|---|---|---|---|
| IC | **D_Dn** | **D_Up** | D | **AmPos** |
| 1 | 1 | Inf | 1 | 0 (< AmPos_Dn) |
| 2 | 1 | Inf | **2.0 - final** | 51 ( $\geq$ AmPos_Dn & $\leq$ AmPos_Up) |

Based on D = 2.0 we have calculated reference parameters $v_i$ according expressions

$$v_i = 1 / \left[ \frac{1}{u_i} + \frac{\log(D)}{T} \right]$$

Final Monte-Carlo simulation was performed with K = 100,000 cycles according reference parameters $v_i$, $i = 1…N$. Final Results after Importance Sampling using (i.e after re-calculations) are following:

P(TOP) = 3.2e-14, STD = 4.9e-16, so Confidence Interval for TOP probability is

[3.0e-14…3.4e-14] with Confidence Level of 0.999

It was also attempted to perform direct Monte-Carlo simulation (i.e. without Importance Samling and re-calculations) of the analysed Fault Tree. Results after performing of the 1,000,000,000 cycles were "zero", i.e. no TOP events were observed. So, although for comparison with proposed algorithm it was used of 10,000 times more Amount of Cysles, results of direct simulation are negative.

## 4. CONCLUSIONS

In this article we have introduced a new algorithm for calculation of the Dynamic Fault Trees. A general purpose Importance Sampling methodology is used for this algorithm development. Main goal was to estimate rare-event Probability of the {TOP = Failure} in an Dynamic Fault Tree having a plurality of Basic Events.
It was assumed, that for each of theBasic Events, the failures are non-repairable and failure times are according general distibution function (Exponential, Weibull, Normal, Log-Normal, etc.). Dynamic Fault Tree may include both Static gates (AND,
OR and based of them composed gates as "K out of M", etc.) and Dynamic gates (PAND, SEQ, SPARE, etc.). The method being performed by the following steps:

a) based on the PDF and CDF for each of the Basic Events, it is constructed a modified, mixed Continious-Discrete, reference PDF.

b) based on this modified reference PDF performing step-by-step preliminary Monte-Carlo simulation of Dynamic Fault Tree untill conditions of optimal reference parameters selection will be satisfyied;

c) selection of the optimal primary reference parameter for each of the Basic Events by means of the optimization under some one common (for all Basic Events) secondary reference parameter D.

d) based on this optimal value of the secondary reference parameter D and corresponding primary reference parameters for each of the Basic Events, performing full Monte-Carlo simulation of the Dynamic Fault Tree and corresponding Importance Sampling re-calculation.

The simulation have gote accurate enough answers and is able to calculate the unavailability for systems which cannot be analytically analyzed.